# Discrimination of low-frequency tones employs temporal fine structure

Tobias Reichenbach and A. J. Hudspeth*

Howard Hughes Medical Institute and Laboratory of Sensory Neuroscience,
The Rockefeller University, New York, NY 10065-6399, USA

*To whom correspondence should be addressed; E-mail: hudspaj@rockefeller.edu.*

**Abstract**

An auditory neuron can preserve the temporal fine structure of a low-frequency tone by phase-locking its response to the stimulus. Apart from sound localization, however, little is known about the role of this temporal information for signal processing in the brain. Through psychoacoustic studies we provide direct evidence that humans employ temporal fine structure to discriminate between frequencies. To this end we construct tones that are based on a single frequency but in which, through the concatenation of wavelets, the phase changes randomly every few cycles. We then test the frequency discrimination of these phase-changing tones, of control tones without phase changes, and of short tones that consist of a single wavelets. For carrier frequencies below a few kilohertz we find that phase changes systematically worsen frequency discrimination. No such effect appears for higher carrier frequencies at which temporal information is not available in the central auditory system.

**Introduction**

In response to a pure tone below 300 Hz, an auditory-nerve fiber fires action potentials at almost every cycle of stimulation and at a fixed phase [1,2]. Above 300 Hz the axon begins to skip cycles, but action potentials still occur at a preferred phase of the stimulus. The quality of this phase locking decays between 1 kHz and 4 kHz, however, and phase locking is lost for still higher frequencies. Phase locking below 4 kHz is sharpened in the auditory brainstem by specialized neurons such as spherical bushy cells that receive input from multiple auditory-nerve fibers [3,4]. These cells can fire action potentials at every cycle of stimulation up to 800 Hz. Temporal information about the stimulus frequency is therefore greatest for frequencies below 800 Hz, declines from 800 Hz to 4 kHz, and vanishes for still greater frequencies.

Phase locking is employed for sound localization in the horizontal plane [5,6]. A sound coming from a subject's left, for example, reaches the left ear first and hence produces a phase delay in the stimulus at the right ear compared to that at the left. Auditory-nerve fibers preserve this phase difference, which is subsequently read out by binaurally sensitive neurons through coincidence detection to determine the angle at which the sound source is located.

The temporal information owing to phase locking might be employed for additional processing of auditory signals in the brain. In particular, phase locking could provide information about the frequency of a pure tone, for the interval between two successive action potentials is on average the signal's period or a multiple thereof. In an accompanying theoretical study we show how neural networks might read out the frequency of a stimulus to high precision [7].

Phase locking has long been hypothesized to aid frequency discrimination [1,2]. For the high frequencies at which temporal fine structure is not preserved in neural responses, the mechanics of the mammalian inner ear spatially separates frequencies sharply enough to account for their discrimination [8,9]. At low frequencies, however, the spatial frequency separation within the cochlea is less pronounced; nevertheless, psychoacoustic experiments show that humans can resolve low frequencies considerably better than high frequencies [8-11]. It is possible that temporal information conveyed through phase locking adds to the spatial frequency information provided by cochlear mechanics. Psychoacoustic experiments on the perception of amplitude- *versus* frequency-modulated tones as well as on complex tones provide indirect evidence for this hypothesis [10,12].

## Results and Discussion

To test directly the usage of temporal information in human frequency discrimination, we constructed tones that are based on a single frequency but in which the phase changes every few cycles. Specifically, we generated wavelets with a carrier frequency $f$ and an amplitude that increases smoothly from zero to a maximal value, remains constant for a certain number of cycles, and eventually returns to zero (Figure 1a). We denote each wavelet's duration, measured in cycles, by $L$. Concatenation of many successive wavelets, in each of which the carrier signal has a random phase, yielded a tone with a random phase change every $L$ cycles (Figure 1a,b). We also generated control tones that have the same amplitude variation as the phase-changing tones but do not exhibit phase changes (Figure 1c).

In the phase-changing tones the information encoded through phase locking is randomly disturbed every $L$ cycles, so the amount of available information corresponds to that in a single wavelet of duration $L$. If phase information is employed in frequency discrimination, then phase-changing tones should be no more differentiable than short tones consisting of only a single wavelet of duration $L$. Frequency discrimination of phase-changing tones should therefore worsen with lower wavelet duration. To test this idea we have also generated short tones that consist of a single wavelet. Because temporal information is not disturbed in the control tones they should allow for much better frequency discrimination that is independent of $L$.

Through psychoacoustic experiments we measured the ability of five normally hearing subjects to discriminate between two close carrier frequencies. For each kind of tone a standard two-interval forced-choice adaptive procedure yielded a threshold value $\Delta f$, the smallest frequency difference that the subject could reliably detect [10] (Figure 2). A lower threshold $\Delta f$ accordingly signifies better frequency discrimination. The dimensionless frequency-difference limen follows as $\Delta f/f$, in which $f$ denotes the average carrier frequency of the presented tones.

We first tested subjects with tones at an average carrier frequency of 500 Hz, a condition in which neuronal responses can be cycle-by-cycle and exhibit phase locking. In all subjects we found that frequency discrimination of both the phase-changing tones and the short tones worsened in a comparable manner when the duration of the wavelets was reduced (Figure 3a). The frequency-difference limen for the phase-changing tones was, however, typically lower than that for short tones of the same duration. This result shows that phase locking is employed for frequency discrimination but is presumably aided by spatial frequency separation in the cochlea. Discrimination of the control tones was always superior and did not depend on the wavelet's duration.

We next performed tests with tones at an average carrier frequency of 5 kHz, a circumstance in which temporal fine structure is not preserved in neural responses. All subjects exhibited similar frequency-difference limens for the phase-changing and the control tones (Figure 3b). The limens did not vary significantly with the duration of the wavelets and were considerably smaller than those for short tones. Evidently no phase information is employed in distinguishing such high-frequency tones.

We finally inquired how the usage of temporal information for frequency discrimination depends on the carrier frequency. To this end we tested the five subjects with tones in which the wavelets had a duration of only seven cycles and varied the carrier frequency between 300 Hz and 5 kHz (Figure 3c). We found that below 1 kHz the phase-changing tones were harder to distinguish than the control tones, whereas above 3 kHz both kinds of tones generally yielded comparable frequency-difference limens. A single subject, **4,** showed worse discrimination of the phase-changing tones up to 5 kHz. In contrast, frequency discrimination of the short tones was typically comparable to that of the phase-changing tones below 1 kHz but worse above 3 kHz. Temporal information is therefore employed below 1 kHz but not much above 3 kHz, in agreement with the presence of phase locking.

The critical frequency at which the frequency discrimination limens for the phase-changing and the control tones became comparable varied from subject to subject. The transition occured at 1 kHz for two subjects (**3** and **5**), at 2 kHz for one subject (**1**), and at 3 kHz for another subject (**2**). The transition point for the remaining subject (**4**) could not be conclusively determined: his discrimination limens for the control tones remained somewhat lower than those for the phase-changing tones up to 5 kHz, although the differences between 2 kHz and 5 kHz were at the edge of statistical significance (see *Statistical analysis* in Materials and Methods). The cycle-by-cycle and phase-locked responses of neurons in the auditory brainstem below about 1 kHz presumably provided superior temporal information that all subjects employed for frequency discrimination. For stimuli of higher frequencies, however, subjects apparently varied in the degree to which they used temporal information.

Temporal information has been assumed to play a role in the appreciation of music as well as in speech recognition [12-14]. The approach that we have developed—quantifying the perception of tones with smooth phase changes through concatenated wavelets—permits testing of the role of phase locking in music and speech processing as well. The results from such experiments might additionally guide the design of future cochlear implants, most of which do not currently evoke phase-locked neural responses [2,15].

## Materials and Methods

### *Ethics statement*

The study was approved by the Institutional Review Board at Rockefeller University under protocol TRE-0748. Written informed consent was obtained from all participants.

### *Sound construction*

A smooth rise in the amplitude $A(t)$ of a wavelet in time $t$ was obtained through the error function:

$$A(t) = \frac{1}{2}\left[\mathrm{erf}\left(\frac{t-t_0}{\delta t}\right)+1\right], \tag{S1}$$

in which $t_0$ denotes the time at which the amplitude has reached half of its maximal value of one and $\delta t$ determines the curve's width, for which we have used two cycles. The decay of the amplitude follows analogously. The wavelet's duration is defined as the number of cycles between the time points at which the amplitude reaches half of its maximal value.

For the phase-changing tones we generated many such wavelets with a carrier frequency $f$ that has a random phase in each wavelet. Through superposition we then concatenated the wavelets such that the amplitude of each had decayed to half of the maximum when the subsequent wavelet's amplitude had risen to the same value. Neither the amplitude nor the phase changed when the carrier waveform had the same phase in both wavelets. If there was a phase change, however, the amplitude of the tone fell transiently because of destructive interference. We concatenated many wavelets to produce tones 0.7 s in duration.

Control tones were obtained by using the envelope of a phase-changing tone to modulate the carrier frequency. There was accordingly no phase change in such a tone. Short tones were individual wavelets.

Because the phase-changing tones resulted from a random sequence of phases in the successive wavelets, we generated ten different realizations for each tone. All tones were computed in Mathematica (Wolfram Research) with a sampling rate of 96 kHz.

### *Stimulus delivery*

A subject seated in a double-walled sound-isolation room (Industrial Acoustics Corporation) viewed a computer monitor outside the room through a double-walled glass window. A computer-generated sound was converted to an analog signal at a sampling rate of 96 kHz by a sound board (M-Audio Audiosport Quattro), amplified by a vacuum-tube amplifier (Stax Systems SRM007t), and delivered to the subject binaurally through electrostatic headphones (Stax Systems SR007a Omega II). The combination of amplifier and headphone had a flat frequency response between 6 Hz and 44 kHz. The phase-changing and control tones were presented at 65 dB SPL. To compensate for the lower audibility of the short tones, which resulted from their brevity, they were delivered at 80 dB SPL.

### *Psychoacoutic testing procedure*

The subjects included two females and three males 26-36 years of age. All subjects except author T. R. were paid for their service.

Subjects interacted with a computer program through a graphical user interface. In each task a subject listened to two successive tones whose carrier frequencies differed by a small amount $\Delta f$: one tone had a carrier frequency that was $\Delta f/2$ above the frequency $f$, and the other tone's frequency was an amount $\Delta f/2$ below. The two tones were separated by a pause of 0.5 s.

The subject was then asked to indicate whether the first or the second tone was lower in frequency. Feedback was provided on the computer monitor, after which the program adapted the frequency difference $\Delta f$ depending on the correctness of the response: three consecutive correct answers resulted in a reduction of the frequency difference whereas a single wrong answer resulted in an increase. The first six changes in frequency difference were by a factor of two and the subsequent ones by a factor of $\sqrt{2}$.

Each subject was trained with all tones until he or she had achieved a stable performance. During an experiment, the first task employed a relatively large frequency difference well above the subject's limen. After an initial phase of ten changes in frequency difference, the subject had settled around an average minimal frequency difference $\Delta f$ (Figure 2). We then presented ten additional changes in frequency difference. The subject's frequency-difference limen and its error were calculated in the logarithmic domain as the average and the standard deviation from the last ten values of $\Delta f$. Because of the adaptive strategy that we employed, each frequency-difference limen corresponded to the frequency difference at which the subject made three successive correct judgments with the same probability as he or she made one incorrect answer, and hence a probability of a correct response of about 70%.

*Statistical analysis*

For each psychoacoustic test we calculated the mean and variance of the frequency-discrimination limen as described above. The mean values and respective standard deviations for the different individuals and different tones are presented in Figure 3. When are the differences between an individual's limens for two types of tones statistically significant? The independent two-sample *t*-test informs us that two observed Gaussian distributions, obtained from ten samples each and with the same standard deviation σ, result from distinct random processes with about 95% probability (*p*-value 0.05) when the means of the two Gaussians differ by 2σ. The probability for distinct underlying processes already exceeds 99% when the two means differ by 3σ. Using a *p*-value of 0.05 as our criterion for statistical significance, we find that two distributions in Figure 3 are distinct if their shaded areas, indicating the standard deviations around the means, do not overlap. Overlapping shaded areas, in contrast, signify a probability of the same underlying stochastic process of more than 5%; we then regard the distributions' differences as not significant.

## Acknowledgments

We thank M. Magnasco for access to a sound-isolation room and electronic equipment, D. Brassil and A. Hurley for help with formulating the clinical protocol, I. Stark and G. Westheimer for discussions, and the members of our research group for comments on the manuscript. This research was supported by grant DC000241 from the National Institutes of Health. T. R. holds a Career Award at the Scientific Interface from the Burroughs Wellcome Fund; A. J. H. is an Investigator of Howard Hughes Medical Institute.

# Figure Legends

**Figure 1. Construction of stimulus sounds.** (**a**) A representative stimulus consists of the concatenation of parts of four wavelets, each ten cycles in duration. Because the carrier frequency has a random phase within each wavelet, the resulting tone displays periodic changes in phase. (**b**) The transition between successive wavelets implies both a phase change and a transient reduction in amplitude. (**c**) A control tone has the same amplitude variation but does not exhibit phase changes.

**Figure 2. Psychoacoustic testing procedure.** The diagram portrays the frequency differences between the two tones in the successive tasks of an exemplary test (black circles). The computer program adapts the frequency difference $\Delta f$ depending on the correctness of the subject's response: the frequency difference is decreased when the subject answers three consecutive tasks correctly whereas a single incorrect answer results in an increase. The average value of $\Delta f/f$ (black line) and its standard deviation (gray shading) are computed from the last ten values of $\Delta f/f$ after the subject's response has reached a steady state.

**Figure 3. Results of psychoacoustic experiments.** The frequency-difference limens of five subjects are presented for phase-changing tones (red squares), control tones (blue triangles), and short tones consisting of single wavelets (black circles). The lines provide a guide to the eye; the shading denotes the standard deviations. (**a**) Frequency-difference limens for tones with a carrier frequency of 500 Hz and different wavelet durations. The limens for the phase-changing and the short tones decrease when the wavelets are lengthened whereas the limens for the control tones remain constant at low values. (**b**) Frequency-difference limens for tones with a carrier frequency of 5 kHz and different wavelet durations. The limens for the phase-changing and the control tones are similar and vary little with the wavelets' duration. The limens for the short tones are significantly greater. (**c**) Frequency-difference limens for wavelets of seven cycles and different

carrier frequencies. The limens for the phase-changing tones exceed those for the control tones below 1 kHz, but the limens begin to converge above 1 kHz.

Reichenbach and Hudspeth

Figure 1

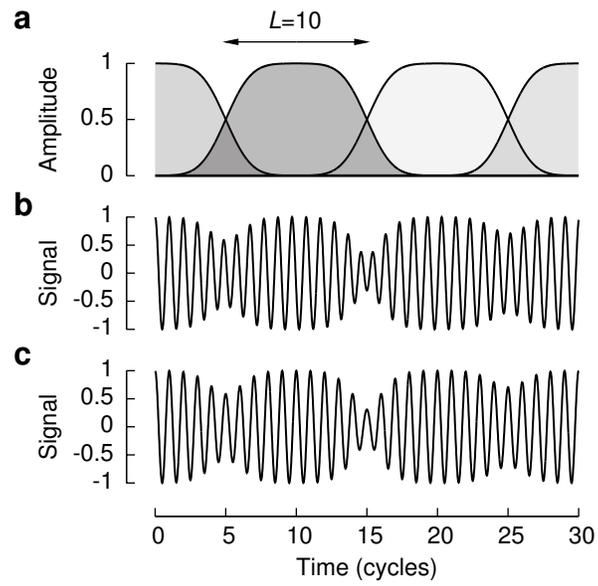

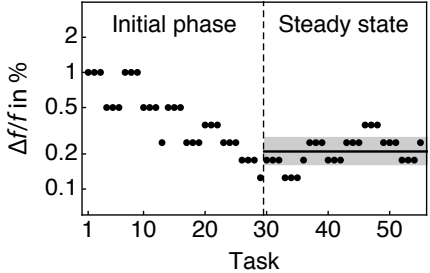

Reichenbach and Hudspeth

Figure 3

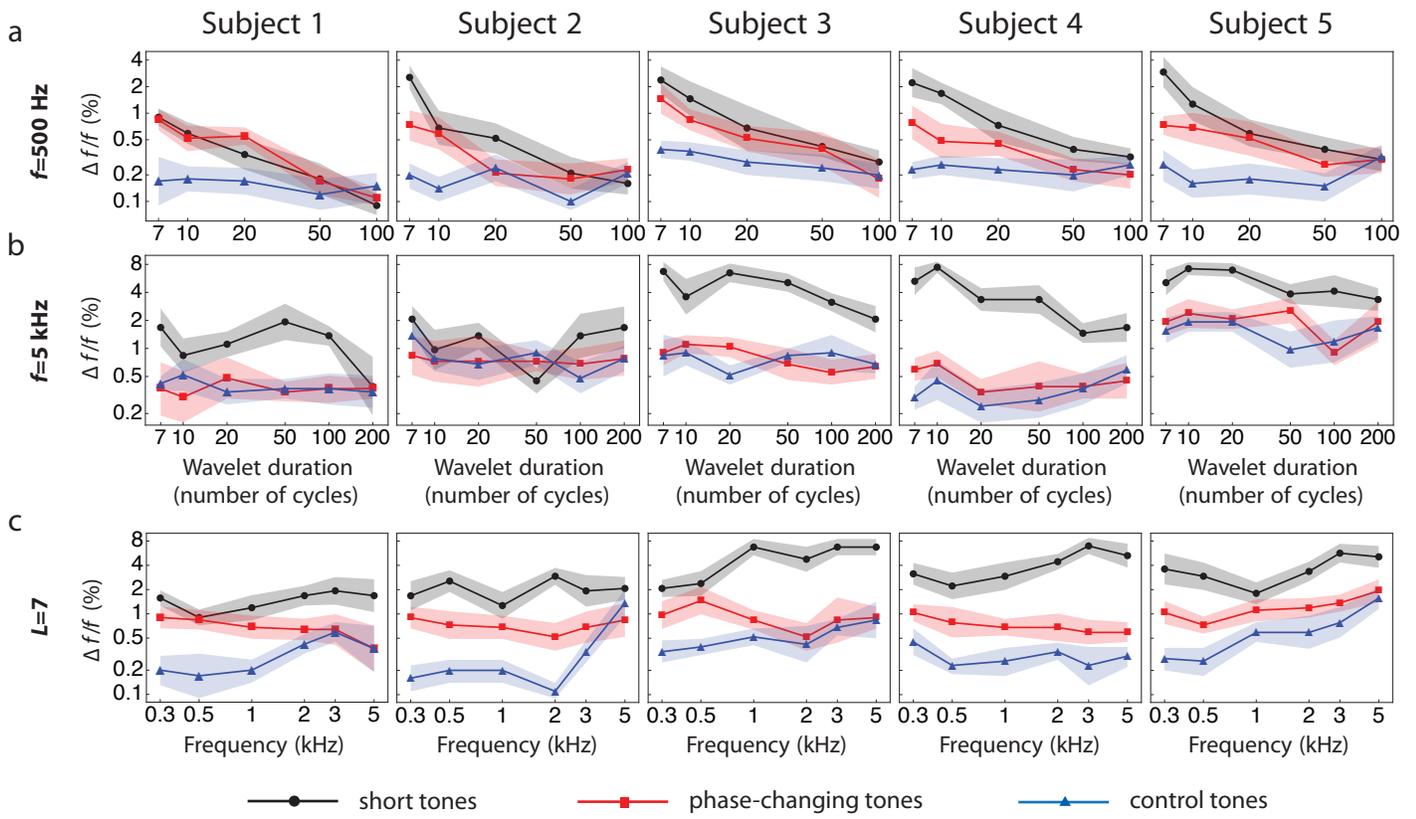